\documentclass[twocolumn,showpacs,prl,aps,groupedaddress]{revtex4-1}
\usepackage{amsmath}
\usepackage{amsfonts}
\usepackage{amssymb}
\usepackage{graphicx}
\usepackage{hyperref}
\usepackage{bm}
\usepackage{color}
\hypersetup{backref,
colorlinks=true,
linkcolor=blue,
linktoc=black,
citecolor=cyan,
urlcolor=black}

\newcommand{\pc}[1]{\ensuremath{\left(#1\right)}}

\begin{document}
\title{
Quasiparticle energy in a strongly interacting homogeneous Bose-Einstein condensate
}

\author{Raphael Lopes}
\email{rl531@cam.ac.uk}
\author{Christoph Eigen}
\author{Adam Barker}
\altaffiliation{Present Address: Clarendon Laboratory, University of Oxford, Oxford OX1 3PU, United Kingdom}
\author{Konrad G. H. Viebahn}
\author{\\Martin Robert-de-Saint-Vincent}
\altaffiliation{Present Address: Laboratoire de physique des lasers, CNRS UMR 7538,
Universit\'e Paris 13, Sorbonne Paris Cit\'e, F-93430, Villetaneuse, France.}
\author{Nir Navon}
\author{Zoran Hadzibabic}
\author{Robert P. Smith}

\affiliation{Cavendish Laboratory, University of Cambridge, J. J. Thomson Avenue, Cambridge CB3 0HE, United Kingdom }

%03.75.Hh	Static properties of condensates; thermodynamical, statistical, and structural properties
%03.75.Nt	Other Bose-Einstein condensation phenomena
%05.30.-d	Quantum statistical mechanics
%67.85.-d 	Ultracold gases, trapped gases
%03.75.Kk 	Dynamic properties of condensates; collective and hydrodynamic excitations, superfluid flow
%42.50.Tx	Optical angular momentum (quantum optics)
%47.37.+q 	Hydrodynamic aspects of superfluidity; quantum fluids
%67.85.De 	Dynamic properties of condensates; excitations, and superfluid flow
%37.10.Vz 	Mechanical effects of light on atoms, molecules, and ions
%37.10.De	Atom cooling methods
%37.10.Gh	Atom traps and guides
%37.10.Jk	Atoms in optical lattices
%67.85.Hj	Bose-Einstein condensates in optical potentials

\begin{abstract}
Using two-photon Bragg spectroscopy, we study the energy of particle-like excitations in a strongly interacting homogeneous Bose-Einstein condensate, and observe 
dramatic deviations from Bogoliubov theory. In particular, at large scattering length $a$ the shift of the excitation resonance from the free-particle energy changes sign from positive to negative. For an excitation with wavenumber $q$, this sign change occurs at $a \approx 4/(\pi q)$, in agreement  with the Feynman energy relation and the static structure factor expressed in terms of the two-body contact. For $a \gtrsim 3/q$ we also see a breakdown of this theory, and better agreement with calculations based on the Wilson operator product expansion. Neither theory explains our observations across all interaction regimes, inviting further theoretical efforts.
\end{abstract}
\maketitle

Spectroscopy of elementary excitations in a many-body system is one of the primary methods for probing the effects of interactions and correlations in the ground state of the system, which are at the heart of macroscopic phenomena such as superfluidity~\cite{kapitza:1938, Allen:1938b}.
In ultracold atomic gases, two-photon Bragg spectroscopy provides a measurement of the excitation energy $\hbar \omega$ at a well defined wavenumber $q$~\cite{Kozuma:1999a,stam99phon,Steinhauer:2002,Papp:2008,Kuhnle:2010,Hoinka2013,Gotlibovych:2014}. 
For a weakly interacting homogeneous Bose-Einstein condensate (BEC), the excitation spectrum is given by the Bogoliubov dispersion relation~\cite{Bogoliubov:1947}, with low-$q$ phonon excitations and high-$q$ particle-like excitations. Predictions of the Bogoliubov theory have been experimentally verified both in harmonically trapped gases, invoking the local density approximation~\cite{stam99phon,Steinhauer:2002}, and in homogeneous atomic BECs~\cite{Gotlibovych:2014}.

Much richer physics, including phenomena traditionally associated with superfluid liquid helium, such as the roton minimum in the excitation spectrum~\cite{Yarnell:1959}, is expected in strongly interacting atomic BECs (for a recent review see~\cite{Chevy:2016}). The strength of two-body interactions, characterised by the s-wave scattering length $a$, can be enhanced by exploiting magnetic Feshbach resonances~\cite{Chin:2010}. However, this also enhances three-body inelastic collisions, making the experiments on strongly interacting bulk BECs~\cite{Papp:2008,Navon:2011,Wild:2012, Makotyn:2014} challenging and still scarce~\footnote{This is in contrast to Fermi gases near Feshbach resonances, and Bose and Fermi gases in optical lattices, where the regime of strong correlations is readily reached experimentally.}. A deviation from the Bogoliubov spectrum was observed in Bragg spectroscopy of large-$q$ excitations in a harmonically trapped $^{85}$Rb BEC \cite{Papp:2008}, and has inspired various theoretical interpretations~\cite{Papp:2008,Ronen:2009,Kinnunen:2009,Sarjonen:2012,Sahlberg:2013,Hofmann:2016,Chevy:2016}, with no consensus or complete quantitative agreement with the experiments being reached so far.

\begin{figure}[t!]
\centering
  \includegraphics[width=\columnwidth]{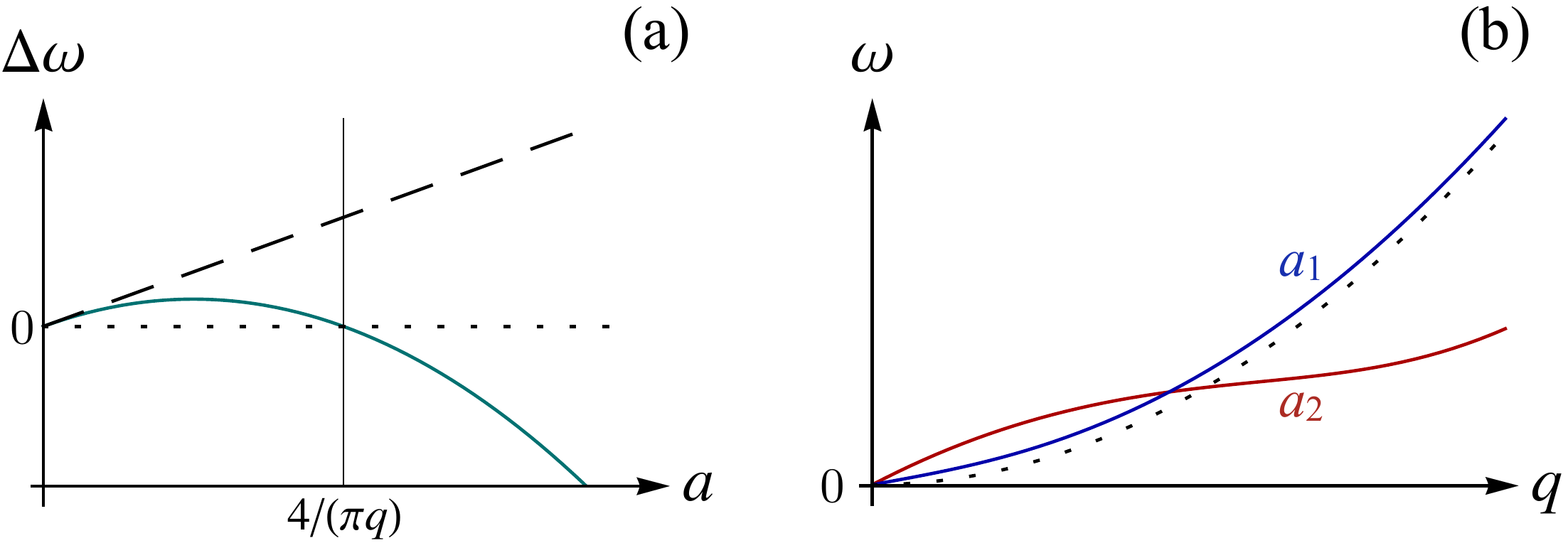}
\caption{(color online) 
Predictions for the excitation resonances. 
(a) Interaction shift for particle-like excitations with a fixed wavenumber $q$. The dashed and solid lines show the Bogoliubov and Feynman-Tan predictions, respectively. 
(b) Sketches of the dispersion relations for two different scattering lengths (solid lines, with $a_2 > a_1$), following~\cite{Rota:2013}. The dotted line shows the free-particle dispersion relation.
}
\label{Fig1}
 \end{figure}

In this Letter, we use Bragg spectroscopy to study the large-$q$, particle-like excitations in a strongly interacting homogeneous $^{39}$K BEC, produced in an optical box trap~\cite{Gaunt:2013}. 
Our homogeneous system allows more direct comparisons with theory, and we also explore stronger interactions than in previous experiments.
We show that at large $a$ the excitation-energy shift from the free-particle dispersion relation strongly deviates from the Bogoliubov theory and even changes sign from positive to negative. 
For $a \lesssim 3/q$ our measurements are in excellent agreement with the calculation based on the Feynman energy relation, with a static structure factor that accounts for short-range two-particle correlations. However, for even stronger interactions we also observe a breakdown of this approximation, and find better agreement with a recent prediction~\cite{Hofmann:2016} based on the Wilson operator product expansion.

In Bogoliubov theory, the  excitation energy $\hbar \omega$ is given by
\begin{equation}
\omega = \omega_0 \sqrt{1 + \frac{2}{q^2\xi^2}} \, ,
\end{equation}
where $\omega_0 = \hbar q^2/(2m)$ is the free-particle dispersion relation, $m$ the atom mass, $\xi = 1/\sqrt{8\pi na}$ the healing length, and $n$ the BEC density.
For particle-like excitations, with $q \gg 1/\xi$, the Bogoliubov prediction for the interaction shift $\Delta \omega = \omega - \omega_0$ is $\Delta \omega_{\rm B} = 4 \pi \hbar n a/m$ [see Fig.~\ref{Fig1}(a)]. This theory assumes $\sqrt{na^3} \ll 1$. Moreover, it is valid only for $q \ll 1/a$, because it does not consider the short-range two-particle correlations, at distances $r\lesssim a$.

For $\sqrt{na^3}\ll 1$,  the Feynman energy relation gives the excitation resonance at $\omega = \omega_0/S(q)$, where $S(q)$ is the static structure factor. Considering short-range correlations, for $q\xi \gg 1$:
\begin{align}
S(q) = 1 + \frac{C}{8n} \pc{\frac{1}{q} - \frac{4}{\pi a q^2}}\, ,
\label{eq:Sq}
\end{align}
where $C (n,a)$ is the two-body contact density, and the expression in the brackets reflects the two-body correlations at short distances~\cite{Pitaevskii:2016,Hofmann:2016}; this `factorisation' of the effects of many-body correlations (captured by $C$) and the short-distance two-body physics was proposed by Tan~\cite{Tan:2008}. For $\sqrt{na^3} \ll1$, the contact density is $C \approx (4\pi na)^2$, and for our experimental parameters $|S(q) - 1| < 0.03$, so $1/S(q) -1  \approx 1 - S(q)$. This `Feynman-Tan' (FT) approach thus gives the interaction shift of the excitation resonance
\begin{align}
  \Delta \omega_{\rm FT} &= \frac{ 4 \pi \hbar na }{m}  \pc{ 1 - \frac{\pi q a}{4}} \, .
\label{eq:Sq1}
\end{align}
For $qa \rightarrow 0$,  $\Delta \omega_{\rm FT}$ reduces to $\Delta \omega_{\rm B}$, but for increasing $a$ (at fixed $q$) it back-bends and changes sign at $a=4/(\pi q)$ [see Fig.~\ref{Fig1}(a)]~\footnote{In~\cite{Papp:2008} the largest value of $a$ reached was $0.8/q$ and back-bending was observed, but $\Delta \omega$ remained positive.}.
At the same time, for the low-$q$ phonons $\Delta \omega$ is positive at all $a$~\cite{Rota:2013}. As illustrated in Fig.~\ref{Fig1}(b), this implies an inflection point in the dispersion relation, $\omega(q)$ at fixed $a$, which is a precursor of the roton minimum that fully develops only for extremely strong interactions~\cite{Rota:2013, Hofmann:2016}. In Eq.~(\ref{eq:Sq}) the maximum in $S(q)$ for fixed $n$ and $a$, which is conceptually associated with the roton~\cite{Steinhauer2005, Hofmann:2016}, occurs at $q = 8/(\pi a)$, independently of $n$, and only for $\sqrt{na^3} \sim 1$ does this coincide with the familiar result for liquid helium, $q_{\rm roton} \sim n^{1/3}$.

In our experiments the regime $\sqrt{na^3} \sim 1$ is not reachable due to significant losses on the timescale necessary to perform high-resolution Bragg spectroscopy. Nevertheless, we reach the regime where interactions are strong enough to observe a dramatic departure from Bogoliubov theory and the precursors of roton physics.
Our setup is described in Ref.~\cite{Eigen:2016}. 
We produce quasi-pure homogeneous $^{39}$K BECs of $N= (50-160) \times 10^3$ atoms in a cylindrical optical box trap of variable radius, $R= (15-30)~\mu$m, and length, $L = (30-50)~\mu$m. The BEC is produced in the 
lowest hyperfine state, which features a Feshbach resonance centred at $402.70(3)$~G~\cite{Fletcher:2017}. By varying $N$, $L$, and $R$, we vary $n$ in the range $ (0.2 - 2.0) \times 10^{12}$~cm$^{-3}$. The three-body loss rate is $\propto n^2 a^4$, so working at such low $n$ is favourable for increasing both $qa$ and $\sqrt{na^3}$.  We prepare the BEC at $a=200~a_0$, where $a_0$ is the Bohr radius, and then ramp $a$ in 50~ms to the value at which we perform the Bragg spectroscopy. For each $n$ we limit $a$ to values for which the particle loss during the whole experiment is $< 10\%$.
By varying the angle between the Bragg laser beams we also explore three different $q$ values: 1.1, 1.7 and 2.0 $k_{\text{rec}}$, where $k_{\text{rec}} = 2\pi /\lambda$ and $\lambda = 767$~nm. For all our parameters we stay in the regime of particle-like excitations, with $q\xi$ values between 5 and 40.

\begin{figure}[t!]
\centering
  \includegraphics[width=\columnwidth]{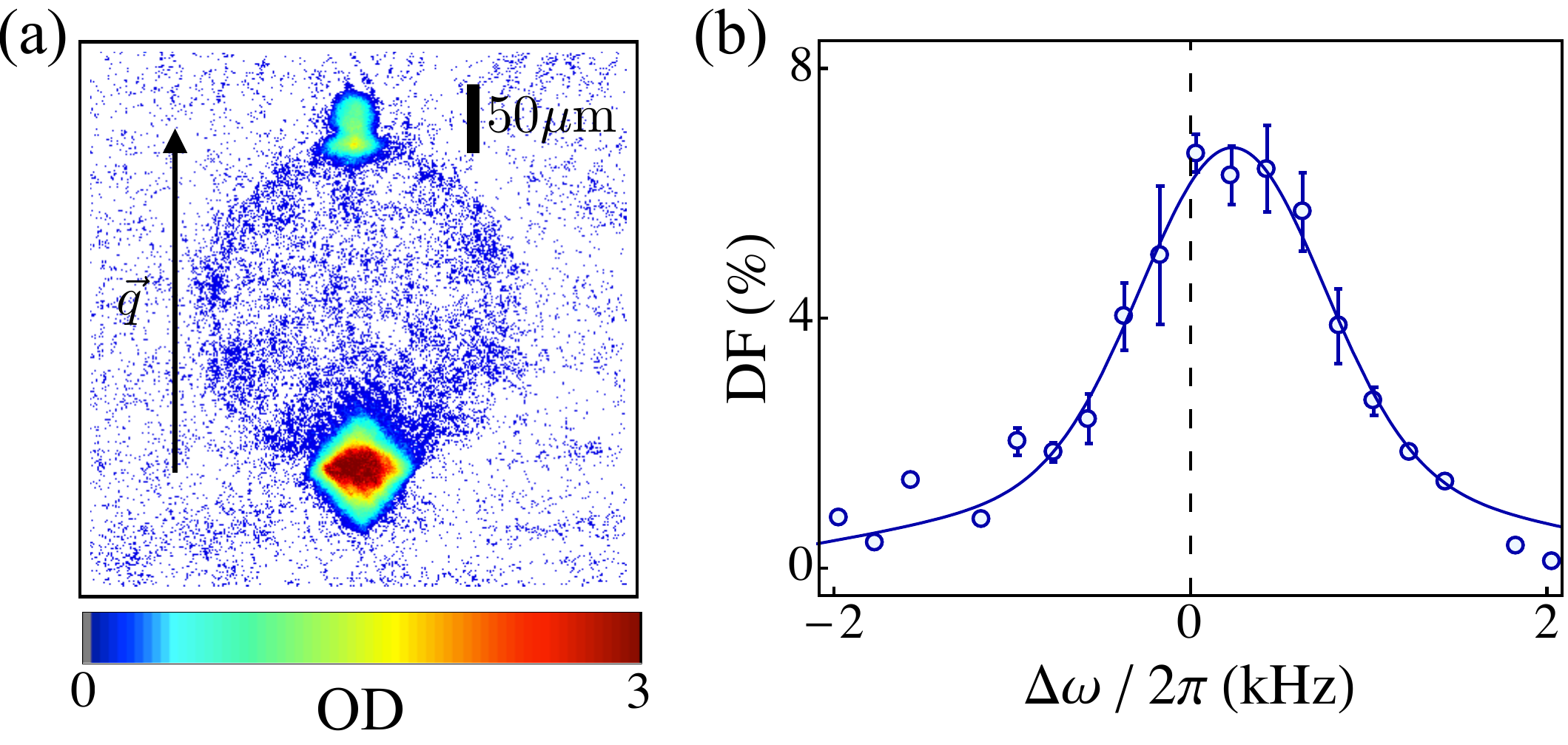}
  \caption{(color online) Bragg spectroscopy, for $n \approx 2.0 \times 10^{12}$~cm$^{-3}$,  $q = 1.7 \, k_{\text{rec}}$, and $a \approx 1000~a_0$. (a) Typical absorption image, taken along the radial direction of the cylindrical box trap, after the 2-ms Bragg pulse and 20~ms of time of flight.
 The spherical halo arises from the collisions between the stationary and diffracted clouds; these collisions do not change the centre of mass of the atomic distribution.
(b) Bragg spectrum. Diffracted fraction (DF) as a function of the frequency difference between the two Bragg beams, referenced to $\omega_0$, which was calibrated using a non-interacting cloud. The resonance is determined from a Gaussian fit to the data (solid line).}
 \label{Fig2}
 \end{figure}

In Fig.~\ref{Fig2}(a) we show an example of an absorption image taken after the Bragg diffraction, and in Fig.~\ref{Fig2}(b) an example of a Bragg spectrum used to determine the resonance shift $\Delta \omega$.  The diffracted fraction of atoms is determined from the centre of mass of the atomic distribution~\cite{Papp:2008, Hoinka2013}; in all our measurements we keep the maximal diffracted fraction to  $\lesssim10\%$.

\begin{figure}[t!]
\centering
  \includegraphics[width=\columnwidth]{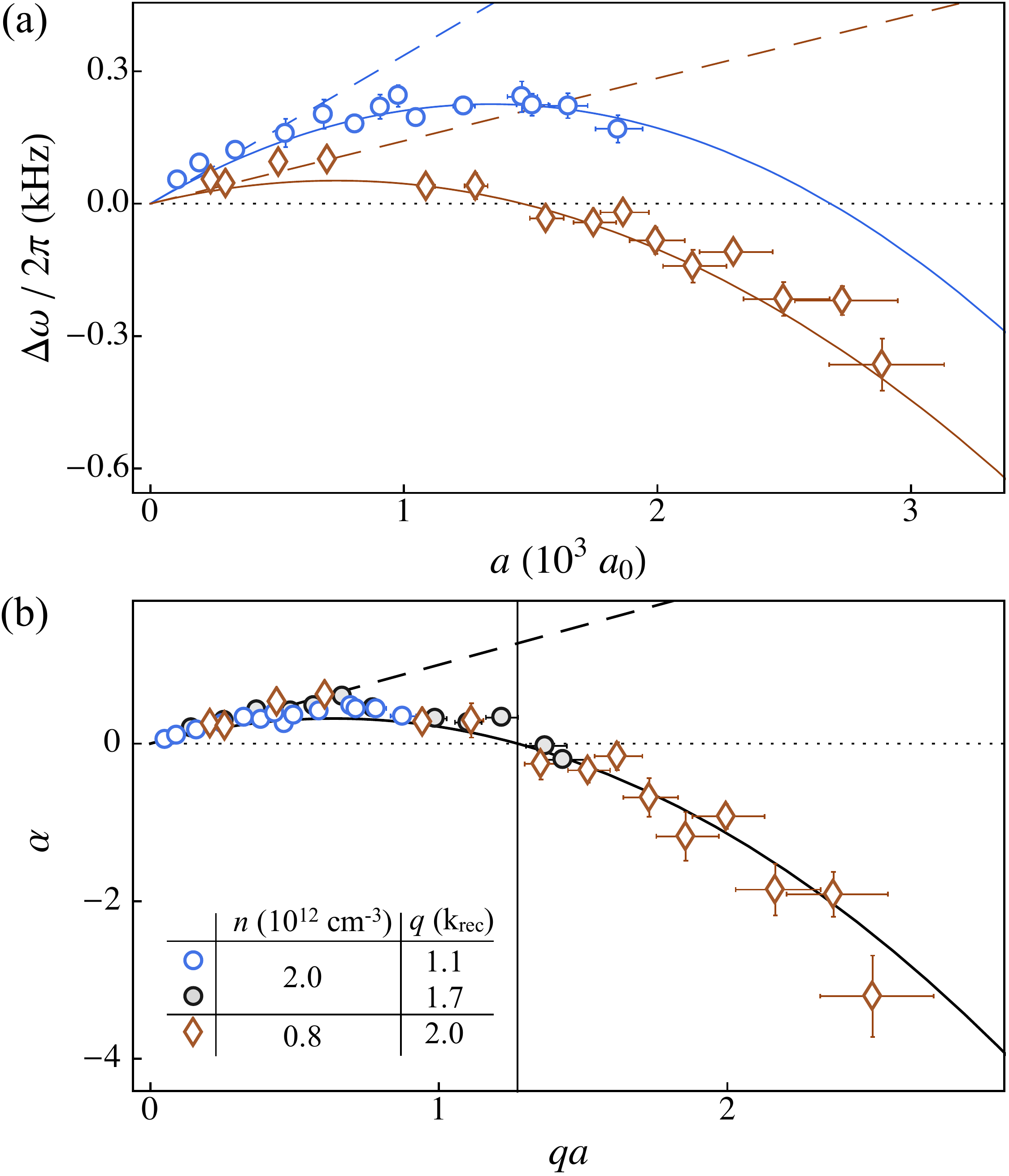}
  \caption{(color online) 
Breakdown of the Bogoliubov approximation and observation of negative frequency shifts.
 (a) $\Delta \omega$ as a function of $a$ for $n \approx 2.0 \times 10^{12}~ \text{cm}^{-3}$ and  $q = 1.1 \, k_{\text{rec}}$ (blue circles), and for $n \approx 0.8 \times 10^{12}~\text{cm}^{-3}$ and $q = 2 \, k_{\text{rec}}$ (orange diamonds).
(b) Dimensionless frequency shift $\alpha$ versus $qa$ for three different combinations of $n$ and $q$. Solid lines in (a) and (b) show the FT predictions from Eqs.~(\ref{eq:Sq1}) and (\ref{eq:Sq2}), respectively, with no adjustable parameters. The dashed lines show the corresponding Bogoliubov predictions.
Vertical error bars show statistical fitting errors and horizontal error bars reflect the uncertainty in the position of the Feshbach resonance.}
\label{Fig3}
 \end{figure}

In Fig.~\ref{Fig3}(a) we plot $\Delta \omega$ versus $a$ for two different combinations of the BEC density $n$ and excitation wavenumber $q$.
In both cases we observe good agreement with the prediction of Eq.~(\ref{eq:Sq1}), without any adjustable parameters; for the lower $n$ we reach higher $a$ and clearly observe that $\Delta \omega$ changes sign.
Defining a dimensionless interaction frequency shift
\begin{align} 
\alpha \equiv \frac{mq}{4\pi\hbar n} \Delta \omega  \, ,
\label{eq:alpha}
\end{align}
the FT prediction of Eq.~(\ref{eq:Sq1}) is recast as:
\begin{align} 
\alpha_{\rm FT} = qa \left(1 - \frac{\pi}{4} qa \right)\, ,
\label{eq:Sq2}
\end{align}
which is a universal function of $qa$ only; with the same normalisation the Bogoliubov theory gives $\alpha_{\rm B} = qa$. 
In Fig~\ref{Fig3}(b) we show that all our measurements of $\alpha$ with three different combinations of $n$ and $q$ fall onto the same universal curve, in agreement with Eq.~(\ref{eq:Sq2})~\footnote{Note that the normalisation in Eq.~(\ref{eq:alpha}) also takes into account the small (10\%) density variations  between measurements taken with different values of $a$ and approximately the same $n$.}.

While in Fig.~\ref{Fig3}(b) all our data agree with Eq.~(\ref{eq:Sq2}),  we note that for the points near $qa=2.5$ the validity of this theory is questionable and the agreement might be partly fortuitous. For these data $\sqrt{na^3} \approx 0.05$, which is already not negligible. At this point the Lee-Huang-Yang (LHY) prediction for the next-order correction to $C$ is of order 50\%~\cite{Lee:1957a,Lee:1957b,Combescot:2009} and even beyond-LHY corrections~\cite{Braaten:1999,Braaten:2002} could be significant. Moreover, the Feynman relation is expected to be quantitatively reliable only for $\sqrt{na^3} \ll 0.1$~\cite{Rota:2013,Rota:2014}. In the final part of the paper we explore even stronger interactions and the limits of validity of the FT prediction.

\begin{figure}[t!]
\centering
\includegraphics[width=\columnwidth]{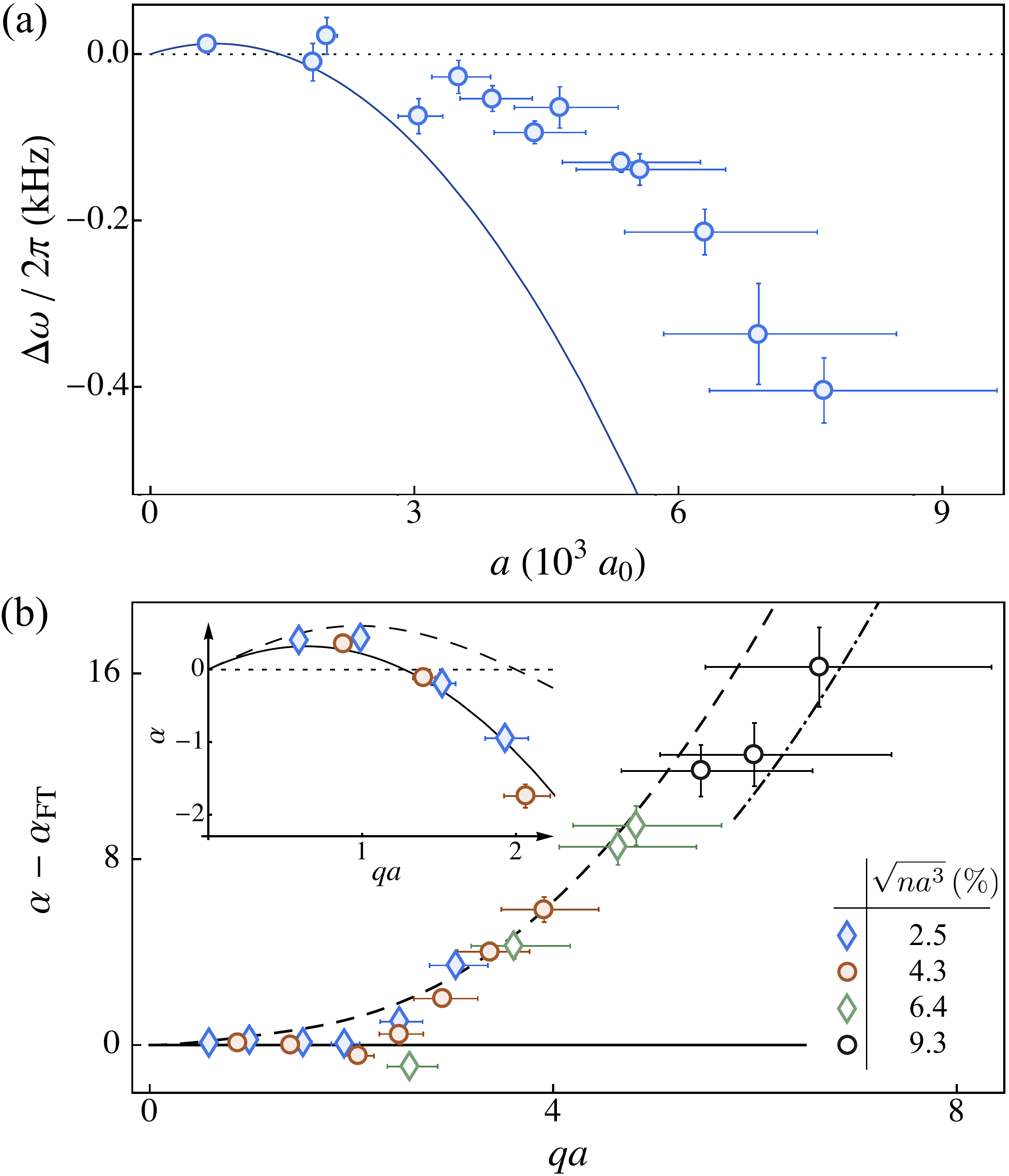}
\caption{(color online) 
Deviation from the Feynman-Tan prediction.
(a)
Frequency shift versus $a$ for $n \approx 0.2 \times 10^{12}~\text{cm}^{-3}$ and  $q = 2\, k_{\text{rec}}$. The solid line shows the FT prediction.
(b) Deviation of the dimensionless frequency shift $\alpha$ from the FT theory as a function of $qa$, for various values of  $\sqrt{na^3}$ (see the legend). The dashed line is the OPE prediction with $C=(4\pi na)^2$ and no adjustable parameters. The dot-dashed line is the OPE prediction that also includes LHY corrections with $\sqrt{na^3} = 0.093$, corresponding to the open-circles data. Inset: comparison of the FT (solid) and OPE (dashed) calculations with the data at low $qa$.
}
\label{Fig4}
 \end{figure}

In Fig.~\ref{Fig4}(a) we show measurements of $\Delta \omega$ with $n \approx 0.2 \times 10^{12}~\text{cm}^{-3}$ and $q = 2 \, k_{\text{rec}}$. Here we explore scattering lengths up to $\approx 8 \times 10^3~a_0$, corresponding to $qa \approx 7$ and $\sqrt{na^3} \approx 0.1$, and observe a strong deviation from the FT prediction.

Tuning $a$ at fixed $n$ and $q$ simultaneously changes $qa$ and $\sqrt{na^3}$, making it non-obvious which of the two dimensionless interaction parameters is (primarily) responsible for the breakdown of the FT theory. In an attempt to disentangle the two effects, we collect data with many $\{n, q, a\}$ combinations, and group them into sets with (approximately) equal $\sqrt{na^3}$, but varying $qa$ values. In Fig.~\ref{Fig4}(b) we plot $\alpha - \alpha_{\rm FT}$ versus $qa$, with different symbols corresponding to different $\sqrt{na^3}$. These measurements strongly suggest that, at least for our range of parameters, the breakdown of the FT theory occurs for $qa \gtrsim 3$, independently of the value of $\sqrt{na^3}$.

At $qa \gtrsim 3$, the deviation of our data from the FT theory is captured well by a recent calculation based on the Wilson operator product expansion (OPE)~ \cite{Hofmann:2016}. Assuming $C=(4\pi na)^2$, 
and with the same normalisation as in Eq.~(\ref{eq:alpha}), 
$\alpha_{\rm OPE} = qa [2/(1 + (qa/2)^2) - 1]$ (see also~\cite{Beliaev:1958});
in Fig.~\ref{Fig4}(b) the dashed black line shows $\alpha_{\rm OPE} - \alpha_{\rm FT}$. 
This theory also allows for self-consistent inclusion of the LHY corrections to $C$, in which case $\alpha_{\rm OPE}$ depends on both  $qa$ and $\sqrt{na^3}$; we show the LHY-corrected $\alpha_{\rm OPE}$ (dot-dashed black line) only for our largest $\sqrt{na^3}$, where it appears to provide a slightly better agreement with the experiments, but this observation is not conclusive (see also~\cite{Wild:2012}).

Finally, we note that while the OPE theory successfully describes our large-$qa$ measurements, it does not agree with our low-$qa$ data, in particular because it predicts the zero-crossing of $\Delta \omega$ at $qa=2$ instead of $qa=4/\pi$; this is highlighted in the inset of Fig.~\ref{Fig4}(b). Providing a unified description of quasiparticle resonances in all interaction regimes thus remains a theoretical challenge.

In conclusion, we have probed the quasiparticle excitations in a strongly interacting homogeneous BEC, pushing the experiments far beyond the regime of validity of the Bogoliubov theory. For a range of interaction strengths ($qa \lesssim 3$), our data can still be quantitatively explained in the framework of the Feynman energy relation, by taking into account the short-range two-particle correlations in the spirit introduced by Tan. However, for our most strongly interacting samples this theory also fails, pointing to the need for more sophisticated theoretical approaches. One such approach, based on the Wilson operator product expansion, accounts well for some of our observations, but does not agree with the experiments in all the interaction regimes that we explored.

We thank Sandro Stringari, Wilhelm Zwerger, Johannes Hofmann, Eric Cornell, Nir Davidson and Stefano Giorgini for inspiring discussions. This work was supported by the Royal Society, EPSRC [Grant No. EP/N011759/1], ERC (QBox), AFOSR, and ARO. R.L acknowledges support from the E.U. Marie-Curie program [Grant No. MSCA-IF-2015 704832] and Churchill College, Cambridge. N.N. acknowledges support from Trinity College, Cambridge.

%\bibliography{CUP_RPSMITH}
%\bibstyle{prsty}


\end{document}